# Magnetic Fields in the Formation of Sun-Like Stars

Josep M. Girart[1*], Ramprasad Rao[2,3], and Daniel P. Marrone[2]

[1]Institut de Ciències de l'Espai (CSIC- IEEC), Campus UAB – Facultat de Ciències, Torre C5 - parell 2ª, Bellaterra, Catalunya 08193, Spain.
[2]Harvard-Smithsonian Center for Astrophysics, 60 Garden St, Cambridge, Massachusetts 02138, USA.
[3]Academia Sinica, Institute of Astronomy & Astrophysics, 645 N. Aohoku Pl., Hilo, Hawaii 96720, USA
[*]To whom correspondence should be addressed: E-mail: girart@ieec.uab.es

**We report high angular resolution measurements of polarized dust emission towards the low-mass protostellar system NGC 1333 IRAS 4A. We show that in this system the observed magnetic field morphology is in agreement with the standard theoretical models of formation of Sun-like stars in magnetized molecular clouds at scales of a few hundred AU; gravity has overcome magnetic support and the magnetic field traces a clear hourglass shape. The magnetic field is substantially more important than turbulence in the evolution of the system and the initial misalignment of the magnetic and spin axes may have been important in the formation of the binary system.**

Magnetic fields are believed to play a crucial role in the formation of stars (1, 2). In the standard model of isolated low-mass star formation (3, 4) magnetized molecular clouds that are magnetically supported against gravitational collapse (or "subcritical") are expected to slowly form dense molecular cores through ambipolar diffusion. The neutral particles are only weakly coupled to the ions, which couple to the magnetic fields, and can drift towards the center of the cloud. The increasing central mass eventually overcomes the magnetic support and the now supercritical core collapses gravitationally. In this collapse phase the initially uniform magnetic field is warped and strengthened in the core. The magnetic field is expected to assume an hourglass shape with an accretion disk formed at the central "pinch" in the field, which corresponds, for an initial cloud size of the order of $10^4$ to $10^5$ AU, to scales of about 100 AU. At large scales, where the contraction effect is small the magnetic field lines are essentially straight. In this axisymmetric scenario the process of magnetic braking, which forces the core to rotate at the same angular speed as the envelope, prevents fragmentation of the collapsing core and the formation of multiple stellar systems. However, models with non-axisymmetric perturbations show that fragmentation can occur on a broad range of scales ($\leq$100 to $10^4$ AU) depending on the initial conditions, such as magnetic field strength, rotation rate, and cloud mass to Jeans mass ratio (5).

The polarization of dust continuum emission provides an opportunity to examine the magnetic field configuration in star forming regions (6). Aspherical spinning dust particles preferentially align themselves with the rotational axis (minor axis) parallel to the direction of the magnetic field. The emission from such grains is partially linearly polarized, with the observed polarization angle perpendicular to the direction of the magnetic field.

NGC 1333 IRAS 4A is a well-studied binary protostellar system in the Perseus molecular cloud complex (7). The two protostars, IRAS 4A1 and A2, are associated with molecular outflows directed roughly north-south (8). The distance to this cloud complex from the Earth is thought to be between 220 and 350 pc (e.g. 9); we adopt the value of 300 pc here. The

Perseus complex is an active star-forming region with around 20 young stellar objects within a projected radius of $4\times10^4$ AU from IRAS 4A. Early polarimetric observations of IRAS 4A (*10,11*) did not have enough angular resolution to be able to examine the spatial scales relevant to the putative hourglass. Yet, in the highest angular resolution observations to date, polarimetry at 230 GHz with the Berkeley-Illinois-Maryland Association (BIMA) array showed hints of an hourglass shape in the magnetic field on 3.5" (1000 AU) scales (*12*).

The Submillimeter Array (SMA) is the first imaging submillimeter interferometer (*13, 14*), providing arcsecond angular resolution and good continuum sensitivity at frequencies that are higher than those currently observable with any other radio interferometer. We used the SMA to observe NGC 1333 IRAS 4A at 345 GHz at an angular resolution of 1.56"×0.99" (with a position angle of 85°). Our data resolve the continuum peaks of IRAS 4A1 and 4A2, which we find to be separated by 400 AU (1.8") at a PA of 130º (see Figure 1), as previously observed at lower frequencies at an angular resolution of ~0.6" (*15*). Using the SMA polarimetry system (*16*) we are able to examine the magnetic field at 360 AU resolution and we find a clear "pinched" morphology (Figure 1C) around this protostellar system. These provide a direct confirmation of the magnetic field configuration at the few hundred AU scale predicted by the standard theory of low mass star formation (*3, 4*). Moreover, the detection of hourglass morphology even in this complex region suggests that the models of isolated star formation may apply even when the initial conditions are much less idealized than is normally assumed. Hints of magnetic field hourglass shape have also been reported in high mass star forming regions such as NGC 2024 (*17*) and more clearly but at much larger scales (~0.5 pc) toward OMC-1 (*18*).

The total flux measured in our 877 μm observations is 6.2±0.5 Jy over an area of 33 square arcsec where there is adequate sensitivity to measure the polarization. Assuming optically thin emission, a dust temperature of 50 K (*19*), a gas to dust ratio of 100, and a dust opacity of 1.5 cm$^{-2}$ g$^{-1}$ (*20*), we estimate the total mass traced by the dust to be 1.2 $d_{300}^2$ M$_\odot$ ($d_{300}\equiv[d/300$ pc$]$, where $d$ is the adopted distance to the NGC 1333 cloud). We can make an estimate of the averaged column and volume density of the region traced by the dust: $<N(H_2)>=M/(A\,\mu_m)$ and $<n(H_2)>=M/(V\,\mu_m)$, where $M$ is the dust mass, $\mu_m$ is the average mass per particle, $A$ is the area of the dust emission, and $V=(4/3)\pi^{-1/2}A^{3/2}$ is the volume. Adopting a helium-to-hydrogen mass ratio of 30%, we find that the mean column density is $<N(H_2)>=8.2\times10^{23}$ cm$^{-2}$ and the mean volume density is: $<n(H_2)>=4.3\times10^7 d_{300}^{-1}$ cm$^{-3}$; both are similar to the expected values for the observed scales (*19*).

With the array configuration and frequency used, these SMA observations are not sensitive to dust emission on scales larger than 10" or 3000 AU where models of magnetized collapsing clouds expect the magnetic field to be uniform. Therefore, the magnetic field has been modelled by a family of parabolic functions using a $\chi^2$ analysis. We find that the center of symmetry of the magnetic field coincides within the measured uncertainty, ~0.6", with the center of the two cores. The position angle of the magnetic field axis, ≈61°, is roughly similar to the orientation of the magnetic field on larger scales around NGC 1333 (*21*). From Figure 1C, we can see that across most of this region there is a remarkably accurate correspondence between the measured magnetic field vectors and the modelled parabolic magnetic field lines. However there are some discrepancies southeast of the centre where the measured field seems to systematically deviate from the fitted model. The observed dispersion (see Fig. 2), $\delta\theta_{obs}$, is made up of contributions from the measurement uncertainty of the polarization angle, $\sigma_\theta$, and the intrinsic dispersion, $\delta\theta_{int}$, according to the equation

(*22*): $\delta\theta_{obs}=(\delta\theta_{int}^2+\sigma_\theta^2)^{1/2}$. The observed dispersion ($\delta\theta_{obs}$) in the residuals is 8.0±0.9°, while the measurement uncertainty of the polarization angle ($\sigma_\theta$) is 6.2±0.3°. Therefore, the intrinsic dispersion is $\delta\theta_{int}$=5.1±1.4°. This estimate of the intrinsic dispersion should be regarded as an upper limit because the parabolic function is just a first approximation of the true magnetic field morphology.

If we assume that the dispersion in polarization angles is a consequence of the perturbation by Alfvén waves or turbulence in the field lines, then the strength of the magnetic field projected in the plane of the sky can be determined from the equation: $B_{pos} = Q (\delta v_{los}/\delta\phi)(4\pi\rho)^{1/2}$. In the previous equation, $\rho$ is the average mass density, $\delta v_{los}$ is the line-of-sight velocity dispersion, $\delta\phi$ is the dispersion in angular deviations of the field lines, which is the same as $\delta\theta_{int}$ calculated above (*23*). $Q$ is a dimensionless parameter that depends on the cloud structure ($Q$=1 corresponds to the original equation of Chandrasekhar & Fermi: ref. *24*). Simulations of turbulent clouds suggest that $Q\approx0.50$ (*25*), which is the value adopted. Using the value of the volume density derived from our data, $n(H_2)=4.3\times10^7$ cm$^{-3}$, and the line width (corrected for the kinematical contribution) given by (*26*), $\delta v_{los}\approx0.2$ km s$^{-1}$, we calculate the magnetic field strength in the plane of the sky to be $B_{pos} \approx 5.0 d_{300}^{-1/2}$ mG.

With this estimate of the magnetic field strength we can compare the properties of NGC1333 IRAS4A derived from our observations to the theoretical predictions. The key parameter that determines whether magnetic fields provide support against gravitational collapse is the mass-to-magnetic flux ratio. Using the formula of (*27*), we find that the mass-to-magnetic flux ratio is $\approx1.7 d_{300}^{1/2}$ times the critical value for collapse. Uncertainties persist due to the neglect of the protostellar mass and the use of the plane-of-sky component of the magnetic field, which respectively increase and decrease this ratio. The estimated mass-to-magnetic flux ratio implies that the region traced by the SMA is slightly supercritical, which is what the theoretical models predict for the observed scales (*4*). This is further supported by the detection of observational signatures of infall motions (*26*).

These data also show that the magnetic energy dominates the turbulent energy in this source. This is demonstrated through $\beta_{turb}$, the square of the ratio of the turbulent line width, $\sigma_{turb}$, to the Alfvén speed, $V_A$. From the expression given by (*17*), we find that $\beta_{turb}$=0.02. Therefore, regardless of whether turbulence played a role in the initiation of the collapse of the parent cloud of NGC 1333 IRAS4A, it seems that at the observed stage of the star formation sequence in this region (Class 0 phase), magnetic fields dominate turbulence as the key parameter to control the star formation process. Finally, the ratio of the magnetic tension to the gravity force, $f_{tension}/f_{gravity}$ demonstrates that gravitational forces are sufficient to cause the observed distortion in the magnetic field. This ratio is proportional to $B^2D^2/(R\rho M)$, where $R$ is the radius of curvature of a given magnetic field line and $D$ is the distance of the origin of this field line to the centre of symmetry (*18*). From the south easternmost-modelled line of Fig. 1, we can estimate $R\approx2.5"$ and $D\approx1.6"$. Using these numbers and the mass derived from the dust emission, we obtain $f_{tension}/f_{gravity} \approx 0.20 d_{300}^{-3}$. This value may be increased by a more accurate model for the magnetic field distribution (which would reduce the residuals and thereby increase the estimated magnetic field) or decreased by including the protostellar mass with the dust mass. Nevertheless it is clear that the two forces are of similar order, as required.

Interestingly, the axis normal to the dusty envelope (44°) lies between the magnetic field axis (61°) and the main outflow axis (19°) (*8*). This suggests that when the collapse initiated, the spin and magnetic axes were not aligned. Could this misalignment be related to the observed formation of a binary system in NGC 1333 IRAS 4A? Studies of collapse in rotating magnetized cores show that fragmentation occurs only in the rotation-dominated cases (when centrifugal forces dominate over magnetic forces) (*28-29*). However, in these cases if the initial spin and magnetic field axes do not coincide the resulting magnetic field direction is expected to be significantly different from its original orientation. This is contrary to the conditions in IRAS 4A where the observed field direction is roughly similar to the larger-scale magnetic field (*19*). In addition, as a consequence of the misalignment the magnetic field geometry is predicted to be considerably distorted from the hourglass shape we observe. The current morphology of this object may indicate that the initial magnetic and centrifugal forces were comparable in magnitude (*29*), allowing fragmentation without significant rotational distortion of the field.

30. The Submillimeter Array is a joint project between the Smithsonian Astrophysical Observatory and the Academia Sinica Institute of Astronomy and Astrophysics and is funded by the Smithsonian Institution and the Academia Sinica. J.M.G. acknowledges Generalitat de Catalunya and Ministerio de Educación y Ciencia (Spain) for support through grants 2004 BE 00370 and AYA2005-08523-C03-02.

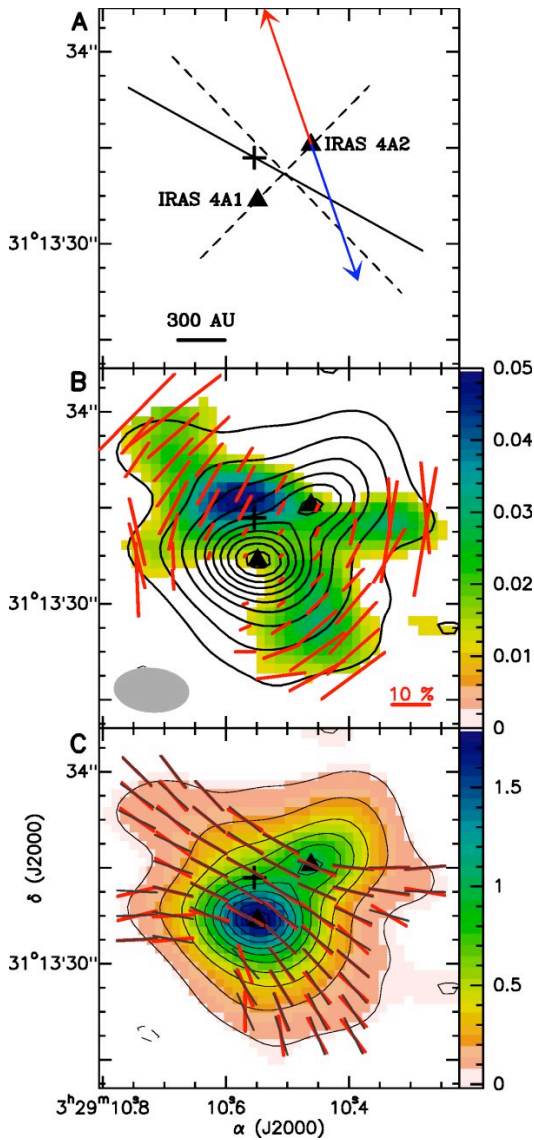

FIGURE 1: (**A**) Sketch of the axis directions: red/blue arrows show the direction of the redshifted/blueshifted lobes of the molecular outflow, probably driven by IRAS 4B (*8*), solid lines show the main axis of the magnetic field, and dashed lines show the envelope axes. The solid triangles show the position of IRAS 4A1 and 4A2. The small cross shows the centre of the magnetic field symmetry. (**B**) Contour map of the 877 μm dust emission (Stokes I) superposed with the color image of the polarized flux intensity. Red vectors: Length is proportional to fractional polarization and the direction is position angle of linear polarization. Contour levels are 1, 3, 6, 9,…30×65 mJy Beam$^{-1}$. The synthesized beam is shown in the bottom left corner. (**C**) Contour and image map of the dust emission. Red bars show the measured magnetic field vectors. Grey bars correspond to the best fit parabolic magnetic field model. The fit parameters are the position angle of the magnetic field axis, $\theta_{PA}=61°\pm6°$, the centre of symmetry of the magnetic field, $\alpha_0(J2000)=3^h29^m10.55^s\pm0.06^s$ and $\delta_0(J2000)=31°13'31.8"\pm0.4"$ and $C=0.12\pm0.06$ for the parabolic form $y=g+gCx^2$, where the *x* is the distance along the magnetic field axis of symmetry from the centre of symmetry.

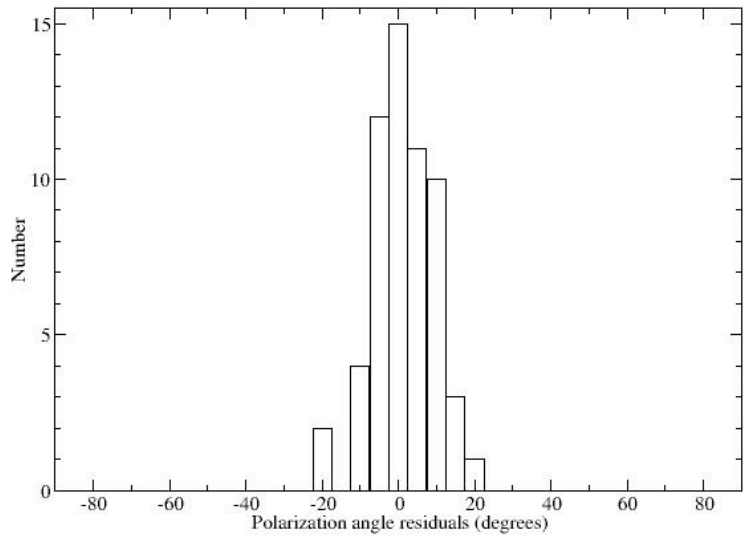

FIGURE 2: Histogram of the polarization angle residuals for the best parabolic magnetic field model, shown in Fig. 1. The mean and the standard deviation of the polarization angle residuals are −1.1° and 8.0°, respectively.